\documentclass[default,iicol]{sn-jnl}% Default with double column layout

\usepackage{graphicx}%
\usepackage{multirow}%
\usepackage{amsmath,amssymb,amsfonts}%
\usepackage{amsthm}%
\usepackage{mathrsfs}%
\usepackage[title]{appendix}%
\usepackage{xcolor}%
\usepackage{textcomp}%
\usepackage{manyfoot}%
\usepackage{booktabs}%
\usepackage{algorithm}%
\usepackage{algorithmicx}%
\usepackage{algpseudocode}%
\usepackage{listings}%
\theoremstyle{thmstyleone}%
\theoremstyle{thmstyletwo}%
\newtheorem{remark}{Remark}%

\theoremstyle{thmstylethree}%

\graphicspath{{./}} %Setting the graphicspath

\usepackage{bm}

\usepackage{lipsum}
\usepackage{mathtools}
\usepackage{cuted}

\begin{document}

\title[Article Title]{Vanquishing volumetric locking in quadratic NURBS-based discretizations of nearly-incompressible linear elasticity: CAS elements}

\author*[1]{\fnm{Hugo} \sur{Casquero}}\email{casquero@umich.edu}

\author[1]{\fnm{Mahmoud} \sur{Golestanian}}

\affil[1]{\orgdiv{Department of Mechanical Engineering}, \orgname{University of Michigan - Dearborn}, \orgaddress{\street{4901 Evergreen Road}, \city{Dearborn}, \postcode{48128-1491}, \state{MI}, \country{U.S.A}}}


\abstract{Quadratic NURBS-based discretizations of the Galerkin method suffer from volumetric locking when applied to nearly-incompressible linear elasticity. Volumetric locking causes not only smaller displacements than expected, but also large-amplitude spurious oscillations of normal stresses. Continuous-assumed-strain (CAS) elements have been recently introduced to remove membrane locking in quadratic NURBS-based discretizations of linear plane curved Kirchhoff rods (Casquero et al., CMAME, 2022). In this work, we propose two generalizations of CAS elements (named CAS1 and CAS2 elements) to overcome volumetric locking in quadratic NURBS-based discretizations of nearly-incompressible linear elasticity. CAS1 elements linearly interpolate the strains at the knots in each direction for the term in the variational form involving the first Lam\'e parameter while CAS2 elements linearly interpolate the dilatational strains at the knots in each direction. For both element types, a displacement vector with $C^1$ continuity across element boundaries results in assumed strains with $C^0$ continuity across element boundaries. In addition, the implementation of the two locking treatments proposed in this work does not require any additional global or element matrix operations such as matrix inversions or matrix multiplications. The locking treatments are applied at the element level and the nonzero pattern of the global stiffness matrix is preserved. The numerical examples solved in this work show that CAS1 and CAS2 elements, using either two or three Gauss-Legrendre quadrature points per direction, are effective locking treatments since they not only result in more accurate displacements for coarse meshes, but also remove the spurious oscillations of normal stresses.}


\maketitle

%\jyear{2021}%

\section{Introduction}\label{sec1}

In commercial finite element analysis (FEA) programs, the use of linear Lagrange polynomials as basis functions dominates in engineering applications involving solid mechanics, particularly in nonlinear dynamic simulations. One of the main reasons why the use of higher-order Lagrange polynomials as basis functions has had very limited success in complex solid mechanics applications is that the higher discrete natural frequencies diverge as the degree of the Lagrange polynomials is increased \cite{Cottrell2006, aleduality, hughes2014finite}. Isogeometric analysis (IGA) \cite{1003.000, cottrell2009isogeometric} proposed the use of nonuniform rational B-splines (NURBS) as basis functions in FEA. In contrast with Lagrange polynomials, even the whole discrete spectrum of natural frequencies may converge to their exact values as the degree of the NURBS basis functions is increased \cite{Cottrell2006, aleduality, hughes2014finite}. In addition, NURBS exhibit enhanced robustness in handling severe mesh distortions in comparison with Lagrange polynomials \cite{Lipton2010}. Thus, the use of NURBS in complex solid mechanics applications has great potential. However, when applied to nearly-incompressible solids or elastoplastic solids, NURBS-based discretizations suffer from volumetric locking as it is also the case for standard FEA discretizations based on Lagrange polynomials \cite{thomas}. Thus, efficient locking treatments for NURBS-based discretizations are needed.

Since our goal is to develop a locking treatment that has the highest possible probability to be adopted in a commercial FEA software, it is useful to first summarize some of the locking treatments that have been widely adopted in commercial FEA software to deal with volumetric locking when using linear Lagrange polynomials. Reduced integration rules have been proposed in which spurious energy modes are handled through the use of hourglass control \cite{flanagan1981uniform, belytschko1984hourglass, reese1999new, reese2000new}. Assumed-strain treatments have been proposed in which the assumed strains are obtained from the compatible strains through either projection or interpolation \cite{nagtegaal1974numerically, hughes1980generalization}. Reduced integration rules with hourglass control and assumed strains are likely the two locking treatments more widely employed by the end users of major FEA programs. Under certain conditions, these two locking treatments are equivalent to mixed methods \cite{hughes1977equivalence, malkus1978mixed, hughes1981general}. Nevertheless, other types of locking treatments are also available in major FEA programs. We mention here the incompatible mode technique \cite{wilson1973incompatible} which was generalized as the enhanced assumed strain (EAS) method using a three-field variational principle \cite{simo1990class, simo1992geometrically}. The original version of the EAS method was later on found to suffer from spurious energy modes in certain cases and new versions of the EAS method were proposed in \cite{glaser1997formulation, kasper2000mixed, kasper2000mixed2} to overcome this issue.

Due to the higher continuity across element boundaries of NURBS, a direct utilization of the numerical schemes developed to overcome volumetric locking in conventional FEA does not effectively remove volumetric locking from NURBS-based discretizations. Thus, new numerical schemes to overcome locking in NURBS-based discretizations are needed. In \cite{auricchio2007fully}, a stream function formulation was developed to remove volumetric locking from NURBS-based discretizations of linear plane-strain problems. In \cite{thomas}, the global $\bar{B}$ projection method was proposed for linear elasticity and the global $\bar{F}$ projection method was proposed for nonlinear elasticity and plasticity. The global $\bar{F}$ projection method was further tested for plasticity in \cite{elguedj2014isogeometric}. The global $\bar{B}$ and $\bar{F}$ are very effective in removing volumetric locking from NURBS-based discretizations of arbitrary degree. However, a significant computational burden is added by these locking treatments. These projection methods are assumed-strain treatments in which the assumed strains are linked to the compatible strains through a $L^2$ projection at the patch level. Thus, these locking treatments require to invert a mass matrix at the patch level and the resulting stiffness matrix at the patch level is not sparse, but full instead. In \cite{antolin2017isogeometric}, an assumed-strain treatment that is inf-sup stable was developed to remove volumetric locking in arbitrary-degree NURBS-based discretizations of nearly-incompressible linear elasticity. This locking treatment is based on a macro-element technique \cite{bressan2011isogeometric, bressan2013isogeometric} which defines the assumed strains on a coarser mesh than the displacement vector, namely, $(p+1)^d$ elements are used for the displacement vector per element used for the assumed strains, where $p$ is the degree of the basis functions used for the displacement vector and $d$ is the number of spatial dimensions. This locking treatment is significantly more computationally efficient than the global $\bar{B}$ method. To the best of the authors' knowledge, it is the only asymptotically-optimal assumed-strain treatment that eliminates volumetric locking from NURBS-based discretizations while having sparse stiffness matrices. However, it still requires matrix inversions and matrix multiplications at the macro-element level to obtain the stiffness matrix and its bandwidth increases with respect to the standard locking-prone NURBS-based discretization of the Galerkin method. In \cite{taylor2011isogeometric}, a three-field mixed method is proposed in which the displacement vector, the mean stress, and the volume effects are independent variables. The volume effect can be condensed out, but the mean stress needs to be kept as an additional unknown. The two-dimensional examples show that the numerical scheme effectively removes volumetric locking for quadratic NURBS-based discretizations of linear elasticity, nonlinear elasticity, and plasticity. In \cite{cardoso2012enhanced}, two EAS methods are proposed for quadratic NURBS-based discretizations of linear plane-strain problems that add four and six additional unknowns. The EAS method that adds four additional unknowns only partially alleviates volumetric locking. Adding this high number of additional unknowns hampers the computational efficiency of the numerical scheme. In \cite{moutsanidis2021reduced, li2022reduced}, a selective integration rule at the element level using Taylor expansions was proposed to alleviate volumetric locking in quadratic NURBS discretizations of linear elasticity, nonlinear elasticity, and plasticity. In \cite{adam2015selective}, reduced and selective integration rules at the patch level are proposed to overcome volumetric locking in linear plane-strain problems using quadratic and cubic NURBS. The selective integration rules are more effective in vanquishing locking than the reduced integration rules since the reduced integration rules need to be slightly over-integrated around the boundary to avoid spurious energy modes. Since the integration rules proposed in \cite{adam2015selective} are at the patch level, a generalization of these integration rules capable of effectively vanquishing volumetric locking in trimmed NURBS \cite{nagy2015numerical, breitenberger2015analysis, leidinger2019explicit, buffa2020minimal, wei2021immersed, antolin2022robust} or in unstructured splines \cite{toshniwal2017smooth, casquero2020seam, hiemstra2020towards, wei2022analysis, shepherd2022isogeometric, shepherd2022feature, toshniwal2022quadratic, wen2023isogeometric} is unlikely to be developed. Locking treatments to vanquish volumetric locking in NURBS-based discretizations that collocate the strong form instead of approximate the variational form have been proposed in \cite{fahrendorf2020mixed, morganti2021isogeometric}.

Continuous-assumed-strain (CAS) elements were recently introduced to remove membrane locking in quadratic NURBS-based discretizations of linear plane curved Kirchhoff rods \cite{casquero2022removing}, to remove shear and membrane locking in quadratic NURBS-based discretizations of linear plane curved Timoshenko rods \cite{casquero2023trods}, and to remove membrane locking in quadratic NURBS-based discretizations of linear Kirchhoff-Love shells \cite{casquero2023klshells}. In this work, we propose two generalizations of CAS elements (named CAS1 and CAS2 elements) that vanquish the volumetric locking existent in quadratic NURBS-based discretizations of linear elasticity combining the following distinctive characteristics: (1) No additional systems of algebraic equations need to be solved, (2) No additional matrix operations such as matrix inversions or matrix multiplications are  needed, (3) No additional unknowns are added, and (4) The nonzero pattern of the stiffness matrix is preserved. To the best of the authors' knowledge, these are the first two assumed-strain locking treatments for quadratic NURBS that combine the aforementioned characteristics. CAS1 elements linearly interpolate the strains at the knots in each direction for the term in the variational form involving the first Lam\'e parameter while CAS2 elements linearly interpolate the dilatational strains at the knots in each direction. When using a displacement vector with $C^1$ continuity across element boundaries (i.e., when no repeated interior knots are introduced), both CAS1 and CAS2 elements result in assumed strains with $C^0$ continuity across element boundaries. Volumetric locking causes not only smaller displacements than expected, but also large-amplitude spurious oscillations of normal stresses. Thus, we study the convergence and plot the distributions of both displacements and stresses to show that CAS1 and CAS2 elements excise the spurious oscillations of normal stresses.

The paper is outlined as follows. Section 2 sets forth the mathematical theory of plane-strain and three-dimensional linear elasticity. Section 3 summarizes how to solve the problem using compatible-strain (CS) elements. Section 4 explains CAS1 and CAS2 elements, the two new element types proposed in this work to overcome volumetric locking in quadratic NURBS-based discretizations. Section 5 evaluates the performance of CS, CAS1, and CAS2 elements with full and reduced integration, including comparisons with exact solutions. Sections 5.1, 5.2, and 5.3 consider the Cook's membrane, an infinite plate with circular hole under in-plane tension, and a three-dimensional block under a compressive volumetric force, respectively. Concluding remarks and directions of future work are drawn in Section 6.

\section{Linear elasticity}\label{sec2}
The domain of the solid is denoted by $\Omega$ and its boundary by $\Gamma$. The displacement vector is denoted by $\bm{u} = \{u_i\}^d_{i=1}$, where $d$ is the number of spatial dimensions. The equations written in this section represent plane-strain linear elasticity for $d=2$ and three-dimensional linear elasticity for $d=3$. In the following, the indices $i,j$, and $k$ take the values $\{1,2,3\}$ for three-dimensional linear elasticity and the values $\{1,2\}$ for plane-strain linear elasticity. Repeated indices imply summation.

\subsection{Infinitesimal strain theory}

The infinitesimal strain tensor is defined as follows
\begin{equation} 
\epsilon_{ij}  = \frac{1}{2} \left(  \frac{\partial u_i}{\partial x_j} + \frac{\partial u_j}{\partial x_i}  \right) \text{.} 
\end{equation}

Note that in the incompressibility limit, $\epsilon_{kk} = \frac{\partial u_k}{\partial x_k} = 0$.

The infinitesimal strain tensor can be additively split into its dilatational (i.e., volumetric) and deviatoric (i.e., isochoric) parts, viz.,
\begin{equation} 
\epsilon_{ij}  = \epsilon^{\text{dil}}_{ij} + \epsilon^{\text{dev}}_{ij} \text{,} 
\end{equation}
with
\begin{align}
   \epsilon^{\text{dil}}_{ij} &=  \frac{1}{d} \frac{\partial u_k}{\partial x_k} \delta_{ij} =  \frac{1}{d} \epsilon_{kk} \delta_{ij} \text{,} \\
   \epsilon^{\text{dev}}_{ij} &=  \frac{1}{2} \left(  \frac{\partial u_i}{\partial x_j} + \frac{\partial u_j}{\partial x_i}  \right) - \frac{1}{d} \frac{\partial u_k}{\partial x_k} \delta_{ij} \text{.} 
\end{align}
where $\delta_{ij}$ is the Kronecker delta, $\epsilon^{\text{dil}}_{ij}$ is the dilatational part of the infinitesimal strain tensor, and $\epsilon^{\text{dev}}_{ij}$ is the deviatoric part of the infinitesimal strain tensor.

\subsection{Linear isotropic material}

For a linear isotropic material, the Cauchy stress tensor has the following expression
\begin{equation} 
\sigma_{ij} = \lambda \epsilon_{kk} \delta_{ij} + 2 \mu \epsilon_{ij} \text{,} 
\end{equation}
where $\lambda$ is the first Lam\'e parameter and $\mu$ is the second Lam\'e parameter. The Lam\'e parameters can be written in terms of the Young's modulus ($E$) and the Poisson's ratio $\nu$ as
\begin{align}
   \lambda&=  \frac{E \nu}{(1+\nu)(1-2\nu)} \text{,} \\
   \mu&=  \frac{E}{2(1+\nu)} \text{.} 
\end{align}
Note how as we approach the incompressibility limit (i.e., as $\nu$ tends to 1/2), $\lambda$ tends to infinity. 

The hydrostatic stress is defined as follows
\begin{equation} 
\sigma_{h} = \frac{\sigma_{kk}}{3} \text{.} 
\end{equation}
In the context of evaluating the performance of numerical schemes with respect to volumetric locking, the hydrostatic stress is a scalar quantity that can be plotted to evaluate whether or not the normal stresses suffer from spurious oscillations. In plane strain, the normal stress in the direction perpendicular to the plane is obtained as 
\begin{equation} 
\sigma_{zz} = \nu (\sigma_{xx} + \sigma_{yy}) \text{.} 
\end{equation}

\subsection{Variational form}

The variational form is obtained from the principle of virtual work. The principle of virtual work states that the internal virtual work ($\delta W^{int}$) must be equal to the external virtual work ($\delta W^{ext}$) for any virtual displacement ($\delta \bm{u}$), i.e.,
\begin{equation} 
 \delta W^{int} =  \delta W^{ext} \quad \forall \delta \mathbf{u} \text{,}  \label{virtualwork}
\end{equation}
with
\begin{align}
   \delta W^{int}&=  \int_{\Omega} \delta\epsilon_{ij}  \sigma_{ij} \, \mathrm d\Omega \nonumber \\
   & = \int_{\Omega} ( \delta\epsilon_{ij} \lambda \epsilon_{kk} \delta_{ij}    + \delta\epsilon_{ij} 2 \mu \epsilon_{ij} ) \, \mathrm d\Omega \text{,}\label{Wint} \\
   \delta W^{ext}&=  \int_{\Omega} \delta u_i  f_i   \, \mathrm d\Omega + \int_{\Gamma_h} \delta u_i h_i    \, \mathrm d\Omega \text{,} \label{Wext}
\end{align}
where $\delta \epsilon_{ij}$ are the virtual strains, $\bm{f} = \{f_i\}^d_{i=1}$ is a force per unit volume, $\Gamma_h$ represents the part of the boundary with Neumann boundary conditions applied, and $\bm{h} = \{h_i\}^d_{i=1}$ is a traction applied (force per unit area). The functional spaces for $\bm{u}$ and $\delta \bm{u}$ are defined taking into account Dirichlet boundary conditions.

\section{Compatible-strain (CS) elements}\label{sec3}

The geometry of the solid is represented as a linear combination of NURBS basis functions, viz.,
\begin{equation}
\bm{x} (\bm{\xi}) = \sum_{A=1}^{n_{cp}} N_A (\bm{\xi}) \bm{Q}_A  \text{,}
\end{equation}
where $\bm{Q}_A$ is the $A$-th control point, $\bm{\xi}$ represents a point in parametric space, and $n_{cp}$ is the total number of control points. In this work, we use open knot vectors with no repeated interior knots and quadratic basis functions. For the details of how to define a geometry using NURBS basis functions and how to perform $h$-refinement using the knot insertion algorithm, the reader is referred to \cite{cottrell2009isogeometric}. Using the isoparametric concept, the displacement vector is discretized as follows
\begin{equation}
\bm{u}^h (\bm{\xi}) = \sum_{A=1}^{n_{cp}} N_A (\bm{\xi}) \bm{U}_A  \text{,}
\end{equation}
where $\bm{U}_A$ is the $A$-th control variable of the displacement vector. Using the Bubnov-Galerkin method, the virtual displacements are discretized as $\delta \bm{u}^h (\bm{\xi}) \in \text{span} \{ N_{A}(\bm{\xi})\}_{A=1}^{n_{cp}}$.

The aforementioned discretization results in the following element stiffness matrix for CS elements
\begin{equation} 
 \mathbf{k} =  \left[ k_{iajb} \right] \text{,} 
\end{equation}
\begin{align} 
 k_{iajb} &=   \int_{\Omega^e} \frac{\partial N_a}{\partial x_i} \lambda \frac{\partial N_b}{\partial x_j}  \, \mathrm d\Omega \nonumber \\  
 & +   \int_{\Omega^e} \frac{\partial N_a}{\partial x_k} \mu \delta_{ij} \frac{\partial N_b}{\partial x_k}  \, \mathrm d\Omega       \nonumber \\  
 & +   \int_{\Omega^e} \frac{\partial N_a}{\partial x_j} \mu  \frac{\partial N_b}{\partial x_i}  \, \mathrm d\Omega          \text{.} 
\end{align}
Following standard FEA paraphernalia, the integrals above are computed performing change of variables twice. First, from the physical coordinates ($\bm{x}$) to the parametric coordinates ($\bm{\xi}$) and then from the parametric coordinates $\bm{\xi}$ to the parent element with coordinates $\widehat{\bm{\xi}} \in [-1,1]^d$. The assembly of the  $n_{el}$ element stiffness matrices into the global stiffness matrix is performed using conventional connectivity arrays as explained in \cite{Hughes2012, cottrell2009isogeometric}, where $n_{el}$ is the total number of elements in the mesh.

\section{Continuous-assumed-strain (CAS) elements}\label{sec4}

The strains of a quadratic CS element have the following expression

\begin{equation} 
\epsilon^{h}_{ij} (\bm{x}) = \frac{1}{2} \left(  \frac{\partial u^{h}_i}{\partial x_j} (\bm{x}) + \frac{\partial u^{h}_j}{\partial x_i} (\bm{x}) \right) \text{.} 
\end{equation}

Thanks to the $C^1$ continuity across element boundaries of the geometry and the displacement vector given by quadratic NURBS, the linear interpolation of the compatible strains at the knots in each direction results in assumed strains with $C^0$ continuity across element boundaries. Thus, the assumed strains of a CAS1 element are defined as follows
\begin{equation} 
\epsilon^{\text{CAS1}}_{ij} (\bm{x}) = \sum_{l=1}^{2^d} L_l (\bm{x}) \epsilon^{h}_{ij} (\bm{x}^e_l)  \text{,} 
\end{equation}
where $\bm{x}^e_l$ are the physical coordinates of the $l$-th corner of element $e$ and $L_l$ is the $l$-th bilinear Lagrange polynomial if $d=2$ (plane-strain linear elasticity) and the $l$-th trilinear Lagrange polynomial if $d=3$ (three-dimensional linear elasticity).

CAS1 elements use the assumed strains for the term of the variational form involving the first Lam\'e parameter and the compatible strains for the terms of the variational form involving the second Lam\'e parameter. Thus, the element stiffness matrix of CAS1 elements is obtained as follows
\begin{equation} 
 \mathbf{k}^{\text{CAS1}} =  \left[ k^{\text{CAS1}}_{iajb} \right] \text{,} 
\end{equation}
%
%\begin{strip}
%\begin{table*}
\begin{align}
& k^{\text{CAS1}}_{iajb} =  \nonumber \\
& \sum_{l=1}^{2^d} \sum_{m=1}^{2^d}   \int_{\Omega^e} L_l (\bm{x}) \frac{\partial N_a}{\partial x_i} (\bm{x}^e_l) \lambda L_m(\bm{x})  \frac{\partial N_b}{\partial x_j} (\bm{x}^e_m)  \, \mathrm d\Omega \nonumber \\  
&  +   \int_{\Omega^e} \frac{\partial N_a}{\partial x_k} \mu \delta_{ij} \frac{\partial N_b}{\partial x_k}  \, \mathrm d\Omega       \nonumber \\  
&  +   \int_{\Omega^e} \frac{\partial N_a}{\partial x_j} \mu  \frac{\partial N_b}{\partial x_i}  \, \mathrm d\Omega          \text{.} 
\end{align}
%\end{table*}
%\end{strip}
%
When using CAS1 elements, the stresses are computed as follows
\begin{equation} 
\sigma^{\text{CAS1}}_{ij} = \lambda \epsilon^{\text{CAS1}}_{kk} \delta_{ij} + 2 \mu \epsilon^{h}_{ij} \text{.} 
\end{equation}
CAS2 elements use the assumed strains for the dilatational part of the infinitesimal strain tensor and the compatible strains for the deviatoric part of the infinitesimal strain tensor, viz.,
\begin{equation} 
\epsilon^{\text{dil, CAS2}}_{ij} (\bm{x}) = \frac{1}{d} \sum_{l=1}^{2^d} L_l (\bm{x}) \epsilon^{h}_{kk} (\bm{x}^e_l) \delta_{ij} \text{,} 
\end{equation}
\begin{equation} 
\epsilon^{\text{dev}, h}_{ij}  = \epsilon^{h}_{ij} - \epsilon^{\text{dil}, h}_{ij} \text{,} 
\end{equation}
\begin{equation} 
\epsilon^{\text{CAS2}}_{ij}  = \epsilon^{\text{dil, CAS2}}_{ij} + \epsilon^{\text{dev}, h}_{ij}  \text{.} 
\end{equation}
Therefore, the element stiffness matrix of CAS2 elements is obtained as follows
\begin{equation} 
 \mathbf{k}^{\text{CAS2}} =  \left[ k^{\text{CAS2}}_{iajb} \right] \text{,} 
\end{equation}
\begin{align} 
 & k^{\text{CAS2}}_{iajb} =  \nonumber \\ 
& \sum_{l=1}^{2^d} \sum_{m=1}^{2^d}   \int_{\Omega^e} L_l (\bm{x}) \frac{\partial N_a}{\partial x_i} (\bm{x}^e_l)  \lambda  L_m(\bm{x})  \frac{\partial N_b}{\partial x_j} (\bm{x}^e_m)  \, \mathrm d\Omega \nonumber \\ 
& + \sum_{l=1}^{2^d} \sum_{m=1}^{2^d}   \int_{\Omega^e} L_l (\bm{x}) \frac{\partial N_a}{\partial x_i} (\bm{x}^e_l)   \frac{2}{d} \mu L_m(\bm{x})  \frac{\partial N_b}{\partial x_j} (\bm{x}^e_m)  \, \mathrm d\Omega \nonumber \\ 
 & -   \int_{\Omega^e} \frac{\partial N_a}{\partial x_i} \frac{2}{d} \mu  \frac{\partial N_b}{\partial x_j}  \, \mathrm d\Omega    \nonumber \\
 & +   \int_{\Omega^e} \frac{\partial N_a}{\partial x_k} \mu \delta_{ij} \frac{\partial N_b}{\partial x_k}  \, \mathrm d\Omega       \nonumber \\  
 & +   \int_{\Omega^e} \frac{\partial N_a}{\partial x_j} \mu  \frac{\partial N_b}{\partial x_i}  \, \mathrm d\Omega          \text{.} 
\end{align}
When using CAS2 elements, the stresses are computed as follows
\begin{equation} 
\sigma^{\text{CAS2}}_{ij} = \lambda \epsilon^{\text{CAS2}}_{kk} \delta_{ij} + 2 \mu \epsilon^{\text{CAS2}}_{ij} \text{.} 
\end{equation}

For both CAS1 and CAS2 elements, the computation of the integrals to obtain each element stiffness matrix and the assembly of each element stiffness matrix into the global stiffness matrix follows the same steps as those summarized for CS elements in the last paragraph of the preceding section. Note that the implementation of CAS1 and CAS2 elements does not require any additional global or element matrix operations such as matrix inversions or matrix multiplications. The locking treatments are applied at the element level and the nonzero pattern of the global stiffness matrix is preserved.

\begin{remark}

If needed to represent a certain geometry exactly, the numerical schemes that define CAS1 and CAS2 elements can be applied to a quadratic NURBS mesh with repeated interior knots, i.e., no changes at all are needed in the numerical schemes to handle a quadratic NURBS mesh with repeated interior knots. In that case, the displacement vector has $C^0$ continuity across the repeated interior knot and the compatible strains are discontinuous across the repeated interior knot. The assumed strains obtained by applying the numerical schemes described in this section are discontinuous across the repeated interior knot. Thus, no matter interior knots are repeated or not, the numerical schemes described in this section lead to assumed strains that preserve the continuity patterns of the compatible strains resulting in an effective locking treatment in both cases. In \cite{casquero2023trods}, the numerical scheme of CAS elements for linear plane Timoshenko rods was applied to quadratic NURBS meshes in which all interior knots are repeated. The resulting element type was called discontinuous-asssumed-strain (DAS) elements since the assumed strains obtained for this type of input mesh are discontinuous across all element boundaries. Comparisons among CAS and DAS elements showed that once locking is properly removed, $C^1$-continuous quadratic NURBS meshes result in higher accuracy than $C^0$-continuous quadratic NURBS meshes. This is the main reason why we focus on $C^1$-continuous quadratic NURBS meshes in this work.

\end{remark}

\begin{remark}

For CAS elements, the dimension of the space used for the displacement vector divided by the dimension of the space for the assumed strains tends to $d$ as the mesh size tends to zero. This is also the case for the global $\bar{B}$ method \cite{thomas, antolin2017isogeometric}.

\end{remark}

%\begin{align} 
% & k^{\text{CAS2}}_{iajb} =  \nonumber \\ 
%& \sum_{l=1}^{2^d} \sum_{m=1}^{2^d}   \int_{\Omega^e} L_l (\bm{x}) N_{a,i} (\bm{x}^e_l) ( \lambda + \frac{2}{3} \mu) L_m(\bm{x})  N_{b,j} (\bm{x}^e_m)  \, \mathrm d\Omega \nonumber \\ 
% & -  \frac{2}{3} \int_{\Omega^e} N_{a,i}  \mu  N_{b,j}  \, \mathrm d\Omega    \nonumber \\
% & +   \int_{\Omega^e} N_{a,k} \mu \delta_{ij} N_{b,k}  \, \mathrm d\Omega       \nonumber \\  
% & +   \int_{\Omega^e} N_{a,j} \mu  N_{b,i}  \, \mathrm d\Omega          \text{.} 
%\end{align}

%\begin{align} 
% k^{\text{CAS1}}_{iajb} &= \sum_{l=1}^{d} \sum_{m=1}^{d}   \int_{\Omega^e} L_l (\bm{x}) N_{a,i} (\bm{x}^e_l) \lambda L_m(\bm{x})  N_{b,j} (\bm{x}^e_m)  \, \mathrm d\Omega \nonumber \\  
% & +   \int_{\Omega^e} N_{a,k} \mu \delta_{ij} N_{b,k}  \, \mathrm d\Omega       \nonumber \\  
% & +   \int_{\Omega^e} N_{a,j} \mu  N_{b,i}  \, \mathrm d\Omega          \text{.} 
%\end{align}

%\begin{align} 
% k^{\text{CAS2}}_{iajb} &= \sum_{l=1}^{d} \sum_{m=1}^{d}   \int_{\Omega^e} L_l (\bm{x}) N_{a,i} (\bm{x}^e_l) ( \lambda + \frac{2}{3} \mu) L_m(\bm{x})  N_{b,j} (\bm{x}^e_m)  \, \mathrm d\Omega \nonumber \\ 
% & -  \frac{2}{3} \int_{\Omega^e} N_{a,i}  \mu  N_{b,j}  \, \mathrm d\Omega    \nonumber \\
% & +   \int_{\Omega^e} N_{a,k} \mu \delta_{ij} N_{b,k}  \, \mathrm d\Omega       \nonumber \\  
% & +   \int_{\Omega^e} N_{a,j} \mu  N_{b,i}  \, \mathrm d\Omega          \text{.} 
%\end{align}

\section{Numerical experiments}\label{sec5}

In this section, we perform numerical investigations using the discretizations explained in Sections 3 and 4. The code used to perform these simulations has been developed on top of the PetIGA framework \cite{dalcin2016petiga}, which adds NURBS discretization capabilities and integration of forms to the scientific library PETSc \cite{petsc-web-page}. Unless mentioned otherwise, a Gauss-Legendre quadrature rule with three integration points per direction is used to compute all the integrals.

Plane-strain problems are known to be particularly prone to volumetric locking. Thus, they constitute demanding benchmark problems to evaluate the performance of locking treatments.

\subsection{Cook's membrane}

\begin{figure}[h!]
\centering
\includegraphics[width=0.8\linewidth]{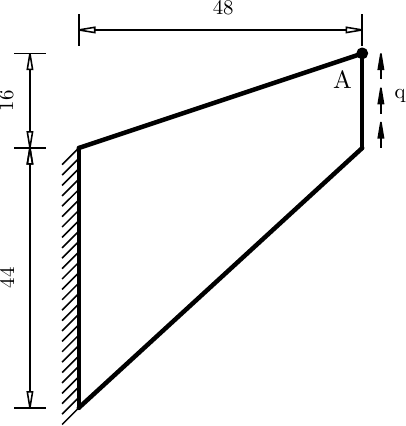}
\caption{Geometry and boundary conditions for the Cook's membrane.}\label{cmgeom}
\end{figure}

\begin{figure}[h!]
\centering
\includegraphics[width=1\linewidth]{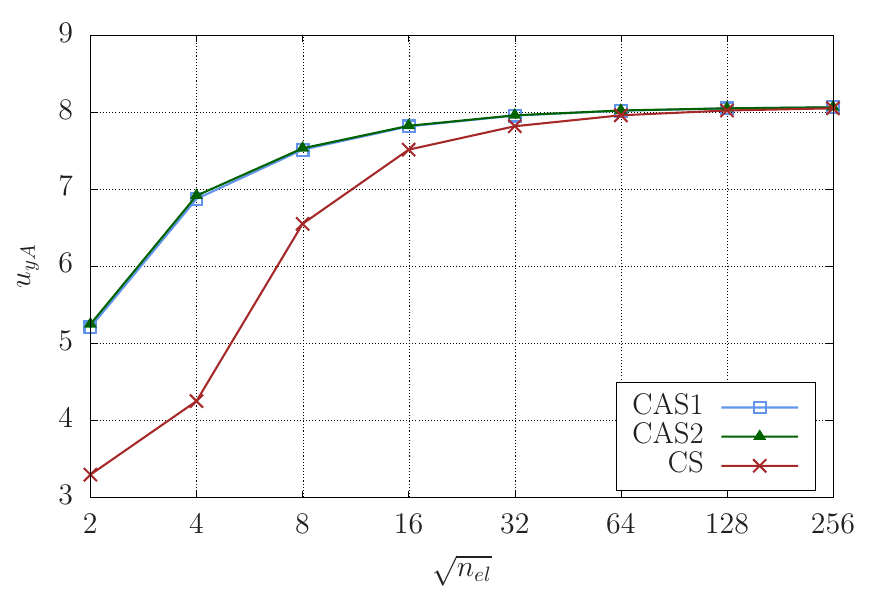}
\caption{(Color online) Cook's membrane. The vertical displacement at the top right corner using CAS1, CAS2, and CS elements is plotted for different mesh resolutions.}\label{cmdisplace}
\end{figure}

\begin{figure*}[h!]
\centering
\includegraphics[scale=1]{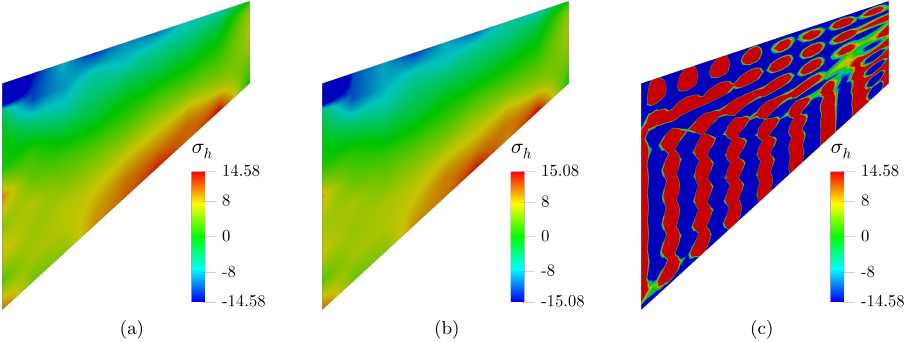}
\caption{Cook's membrane. (a) Distribution of the hydrostatic stress using CAS1 elements. (b) Distribution of the hydrostatic stress using CAS2 elements. (c) Distribution of the hydrostatic stress using CS elements. A mesh with $8 \times 8$ elements is used in all three plots.}\label{cmsigmah}
\end{figure*}

The first problem is a widespread benchmark problem to evaluate the performance of discretizations of nearly-incompressible solids under combined bending and shear \cite{cook2007concepts, cesar1999new, kasper2000mixed, chavan2007locking, moutsanidis2021reduced, thomas}. It consists of a parallelogram that is fixed at one side and has a tangential force per unit area applied at the opposite side whose value is $q = 6.25$. Homogeneous Neumann boundary conditions are applied at the other two sides. The geometry and the boundary conditions of this problem are shown in Fig. \ref{cmgeom}. The problem is solved under the assumption of plain strain. The Young's modulus and Poisson's ratio used in this problem are
\begin{equation} 
 E =  240.565,  \quad \nu = 0.4999 \text{,} 
\end{equation}
respectively. The reference value for the vertical displacement at point A (point A is indicated in Fig. \ref{cmgeom}) is 8.075. We initiate our convergence study with a uniform mesh composed of $ 2 \times 2$ $C^1$-continuous quadratic elements. Subsequently, we carry out uniform $h$-refinement seven times using the knot insertion algorithm. Fig. \ref{cmdisplace} plots the convergence of the vertical displacement at point A using CS, CAS1, and CAS2 elements. As shown in Fig. \ref{cmdisplace}, CS elements need significantly more elements than CAS1 and CAS2 elements to obtain an accurate result. For a given mesh, the displacement values obtained with CAS1 and CAS2 elements are indistinguishable at the scale of the plot.

Fig. \ref{cmsigmah} plots the distribution of the hydrostatic stress using $ 8 \times 8$ CAS1 elements, $ 8 \times 8$ CAS2 elements, and $ 8 \times 8$ CS elements. As discussed in \cite{schroder2021selection}, this benchmark problem has a singularity at the top left corner. Thus, in Figs. \ref{cmsigmah} a) and b), we use scales for the hydrostatic stress whose minimum value has the same absolute value as the maximum hydrostatic stress of the numerical solution, but the opposite sign. As shown in Fig. \ref{cmsigmah}, CAS1 and CAS2 elements are free from spurious oscillations. However, CS elements have large-amplitude spurious oscillations. In Fig. \ref{cmsigmah} c), we use the same scale for CS elements as for CAS1 elements to better show that the number of spurious oscillations is related to the number of elements in the mesh. This was also the case for rods \cite{casquero2022removing, casquero2023trods}. The errors in stresses will be properly quantified in the benchmark problem considered next by leveraging the fact that its exact solution is known.

\subsection{Infinite plate with a circular hole under in-plane tension}

\begin{figure}[h!]
\centering
\includegraphics[width=0.9\linewidth]{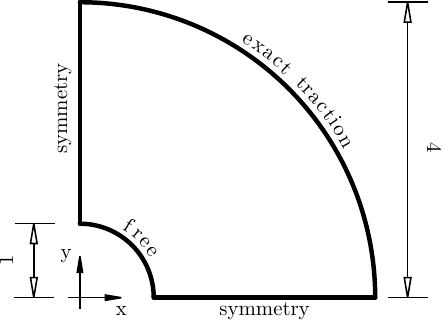}
\caption{Geometry and boundary conditions for the infinite plate with a hole under tension.}\label{ipgeom}
\end{figure}

The second problem is a infinite plate with a circular hole under uniaxial tension in the horizontal direction. This problem has been used in preceding works to evaluate the performance of discretizations of nearly-incompressible solids since its exact solution is known \cite{dolbow1999volumetric, huerta2001locking, thomas, adam2015selective}. Under the assumption of plain strain, the exact solution of this problem is

\begin{align}
u_{x}(r,\theta)& = \frac{T_xR}{8 \mu} \Biggl[ \left( 4-4 \nu \right) \frac{r}{R} \cos \left( \theta \right)  \Biggr. \nonumber \\
& + \frac{2R}{r} \left( \left( 4-4 \nu \right) \cos \left( \theta \right) + \cos \left( 3 \theta \right) \right) \nonumber \\
& \Biggl. -2\frac{R^3}{r^3} \cos \left( 3 \theta \right) \Biggr] , \\
u_{y}(r,\theta)& = \frac{T_xR}{8 \mu} \Biggl[ \left( -4 \nu \right) \frac{r}{R} \sin \left( \theta \right)  \Biggr. \nonumber \\
& + \frac{2R}{r} \left( \left( 4 \nu - 2 \right) \sin \left( \theta \right) + \sin \left( 3 \theta \right) \right) \nonumber \\
& \Biggl. -2\frac{R^3}{r^3} \sin \left( 3 \theta \right) \Biggr] ,
\end{align}

\begin{align}
\sigma_{x x} =T_x &\left( 1-\frac{R^2}{r^2} \left( \frac{3}{2} \cos (2 \theta)+\cos (4 \theta ) \right) \right. \nonumber \\
& \left. +\frac{3 R^4}{2 r^4} \cos (4 \theta)\right) \label{sigmaxxexact}\text{,} \\
\sigma_{y y} =T_x &\left(-\frac{R^2}{r^2}\left(\frac{1}{2} \cos (2 \theta)-\cos (4 \theta)\right) \right. \nonumber \\
& \left. -\frac{3 R^4}{2 r^4} \cos (4 \theta)\right) \text{,} \\
\sigma_{x y} =T_x &\left(-\frac{R^2}{r^2}\left(\frac{1}{2} \sin (2 \theta)+\sin (4 \theta)\right) \right. \nonumber \\
& \left. +\frac{3 R^4}{2 r^4} \sin (4 \theta)\right) \text{,} 
\end{align}

with

\begin{align}
r &= \sqrt{\left(x^2+y^2\right)^2}  \text{,} \\
\theta &= \tan ^{-1}\left(\frac{y}{x}\right) \text{,} 
\end{align}

where $T_x$ is the uniaxial horizontal tension applied and $R$ is the radius of the hole.

\begin{figure}[h!]
\centering
6\includegraphics[width=1\linewidth]{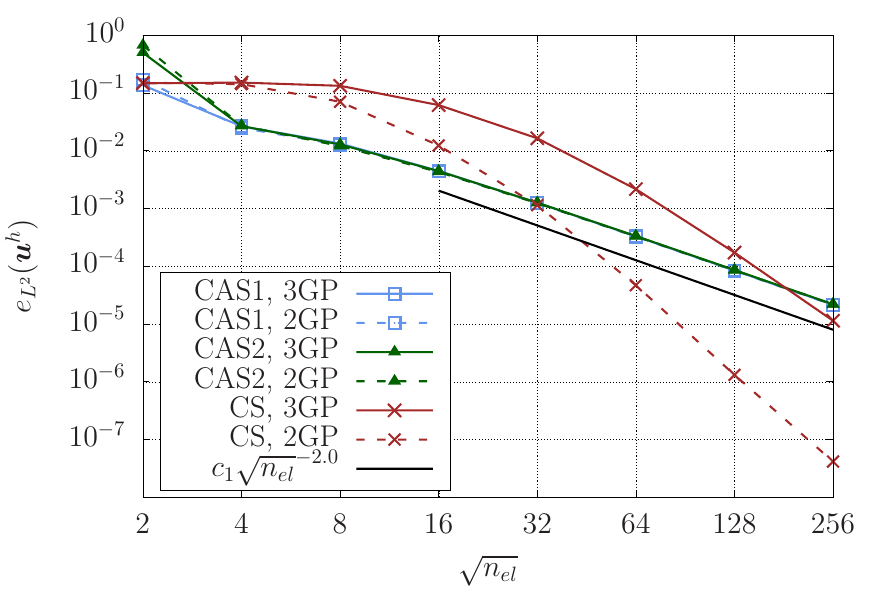}
\caption{(Color online) Infinite plate with a hole under tension. Convergence of the displacement vector in $L^2$ norm. At the scale of the plot, the numerical solution using CAS1 and CAS2 elements overlap using either two or three Gauss-Legendre quadrature points per direction.}\label{ipdisp}
\end{figure}

\begin{figure}[h!]
\centering
6\includegraphics[width=1\linewidth]{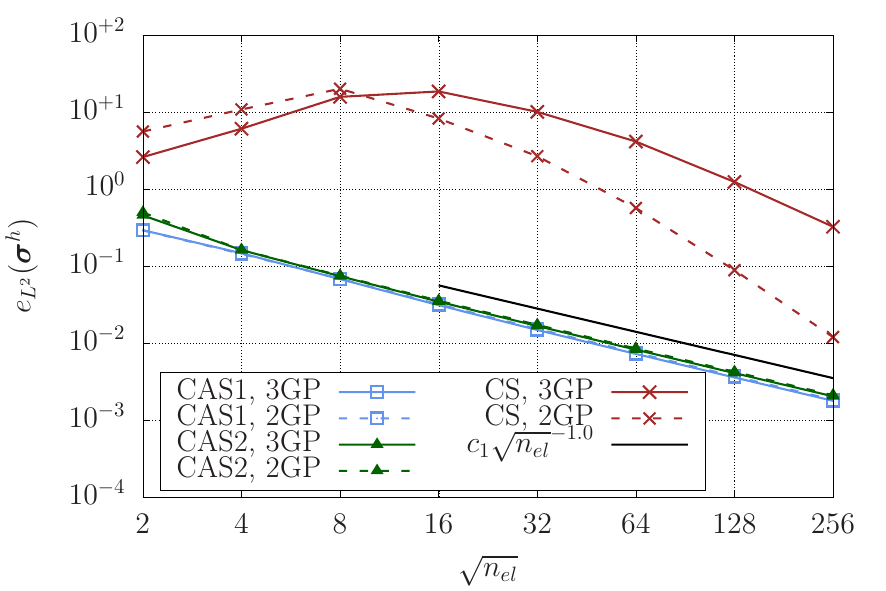}
\caption{(Color online) Infinite plate with a hole under tension. Convergence of the stress tensor in $L^2$ norm. At the scale of the plot, the numerical solution using CAS1 and CAS2 elements overlap using either two or three Gauss-Legendre quadrature points per direction.}\label{ipstress}
\end{figure}

\begin{figure*}[h!]
\centering
\includegraphics[scale=1]{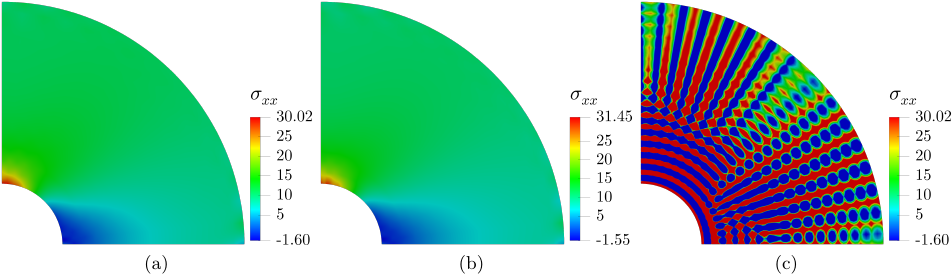}
\caption{Infinite plate with a hole under tension. (a) Distribution of the normal stress in $x$ direction using CAS1 elements. (b) Distribution of the normal stress in $x$ direction using CAS2 elements. (c) Distribution of the normal stress in $x$ direction using CS elements. A mesh with $16 \times 16$ elements is used in all three plots.}\label{iphstress}
\end{figure*}

Due to the symmetry of the problem, only a quarter of the geometry is considered in the simulations. In addition, the infinite plate needs to be approximated by a finite plate in the simulations. We choose to consider a circular plate with an outer radius whose value is 4. The exact values of the stresses are evaluated at the outer radius of our circular plate and enforced as a Neumann boundary condition. The geometry and the boundary conditions of this problem are shown in Fig. \ref{ipgeom}. The Young's modulus and Poisson's ratio used in this problem are
\begin{equation} 
 E =  1 \times 10^5,  \quad \nu = 0.49999 \text{,} 
\end{equation}
respectively. We initiate our convergence study with a uniform mesh composed of $ 2 \times 2$ $C^1$-continuous quadratic elements. The geometry is represented exactly since we are using quadratic NURBS. After that, we carry out uniform $h$-refinement seven times using the knot insertion algorithm. We use the exact solution of this problem to study the convergence in $L^2$ norm of the displacements and the stresses. In order to do so, we define the relative errors in $L^2$ norm of the displacement vector and the Cauchy stress tensor as
\begin{align}
   e_{L^2}(\mathbf{u}^h) &=  \frac{ \sqrt{ \int_{\Omega}  \sum_{i=1}^{2} \left( u^h_i - u_i \right)^2 \, \mathrm d\Omega } }{ \sqrt{ \int_{\Omega} \sum_{i=1}^{2}  u^2_i \, \mathrm d\Omega  } }  \text{,} \\
   e_{L^2}(\bm{\sigma}^h) &=  \frac{ \sqrt{\int_{\Omega} \sum_{i=1}^{2} \sum_{j=1}^{2} \left( \sigma^h_{ij} - \sigma_{ij} \right)^2 \, \mathrm d\Omega }}{ \sqrt{ \int_{\Omega} \sum_{i=1}^{2} \sum_{j=1}^{2} \sigma^2_{ij} \, \mathrm d\Omega} } \text{.}
\end{align}
Since we are solving second-order partial differential equations with basis functions of degree 2, the optimal asymptotic convergence rates of $e_{L^2}(\mathbf{u}^h)$ and $e_{L^2}(\bm{\sigma}^h)$ are 3 and 2, respectively \cite{Hughes2012}. Figs. \ref{ipdisp} and \ref{ipstress} plot the convergence in $L^2$ norm of the displacements and the stresses, respectively, using CS elements, CAS1 elements, and CAS2 elements with both two and three Gauss-Legendre quadrature points per direction. As shown in Figs. \ref{ipdisp} and \ref{ipstress}, the convergence curves of CS elements, with either two or three Gauss-Legendre quadrature points per direction, suffer heavily from locking. Specifically, Fig. \ref{ipstress} shows how the relative error in $L^2$ norm of the stresses increases as uniform $h$-refinement is performed multiple times. In addition, for quite refined meshes, the relative error in $L^2$ norm of the stresses can be still high (namely, greater than 100\%) while the relative error in $L^2$ norm of the displacements is already low (namely, smaller than 1\%). This fact evidences the need for studying the accuracy of both the displacements and the stresses when evaluating the performance of a numerical scheme to deal with volumetric locking. In contrast, as shown in Figs. \ref{ipdisp} and \ref{ipstress}, CAS1 and CAS2 elements do not suffer from locking since their convergence curves have a clear convergence pattern throughout the eight mesh resolutions considered. For CAS1 and CAS2 elements, the same level of accuracy is obtained using either two or three Gauss-Legendre quadrature points per direction. Thus, employing two Gauss-Legendre quadrature points per direction is preferable to decrease the computational time spent in computing each element stiffness matrix. For most engineering applications, discretization errors of 1\% are acceptable since such errors are likely to be smaller than model errors (errors between reality and the mathematical model). CS elements have optimal asymptotic convergence rates while CAS1 and CAS2 elements do not. However, for either the displacements or the stresses, CAS1 and CAS2 elements obtain relative errors in $L^2$ norm smaller than 1\% for significantly coarser meshes than CS elements (particularly for the stresses). This evidences that optimal asymptotic convergence rates are not the only metrics that one should focus on when evaluating the suitability of a numerical scheme for engineering applications.

Fig. \ref{iphstress} plots the distribution of the normal stress in $x$ direction using $16 \times 16$ CAS1 elements, $16 \times 16$ CAS2 elements, and $16 \times 16$ CS elements. Using Eq. \eqref{sigmaxxexact}, we obtain that the maximum value of the exact solution of the normal stress in $x$ direction for this problem is 30. As shown in Fig. \ref{iphstress}, CAS1 and CAS2 elements are free from spurious oscillations. However, CS elements have spurious oscillations. In Fig. \ref{iphstress} c), we use the same scale for CS elements as for CAS1 elements to better show that once again the number of spurious oscillations is related to the number of elements in the mesh.

\subsection{Three-dimensional block under a compressive volumetric force}

\begin{figure}[h!]
\centering
\includegraphics[width=0.8\linewidth]{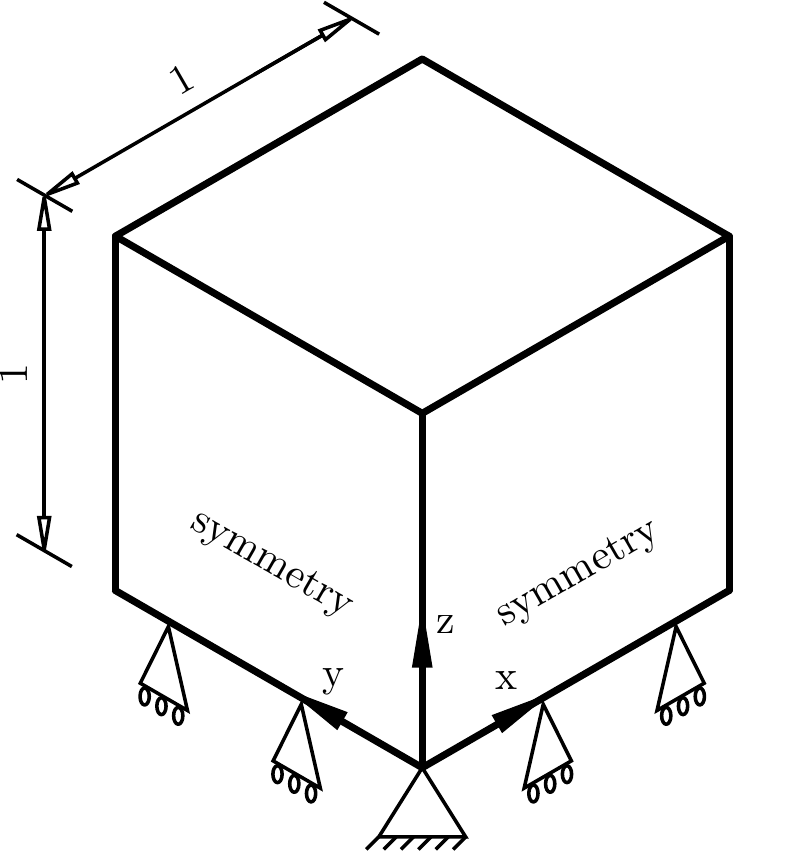}
\caption{Geometry and boundary conditions for the three-dimensional block under a compressive volumetric force.}\label{3dgeom}
\end{figure}

\begin{figure*}[h!]
\centering
\includegraphics[scale=1]{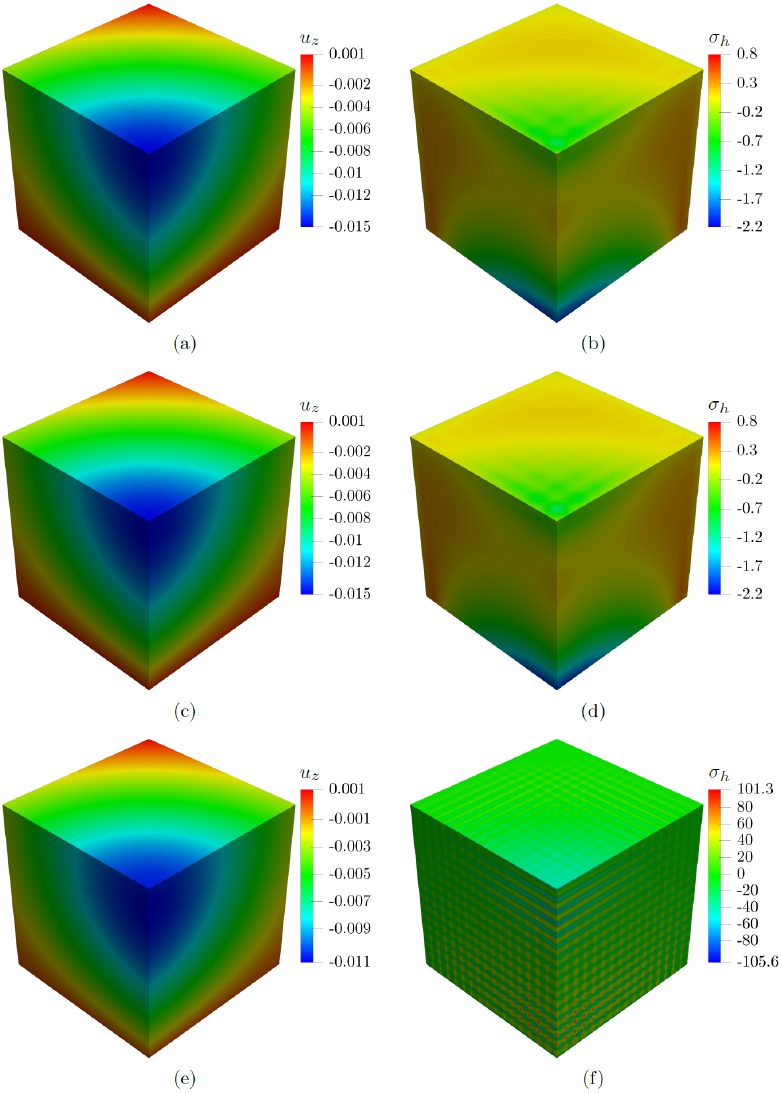}
\caption{Three-dimensional block under a compressive volumetric force. (a), (c), and (e) plot the vertical displacement using CAS1, CAS2, and CS elements, respectively. (b), (d), and (f) plot the hydrostatic stress using CAS1, CAS2, and CS elements, respectively. A mesh with $16 \times 16 \times 16$ elements is used in all the plots.}\label{3dplots}
\end{figure*}

The last example consists of a three-dimensional block with zero vertical displacements at the bottom plane and undergoing compression through the application of a force per unit volume. Due to the symmetry of the problem, only a quarter of the geometry is considered in the simulations. The geometry and the boundary conditions of this problem are shown in Fig. \ref{3dgeom}. The force per unit of volume has the following expression
\begin{align}
  f_x &= 0  \text{,} \\
    f_y &= 0  \text{,} \\
   f_z &= - 10 ( 1 - \lvert x \rvert ) (1 - \lvert y \rvert) \text{,}
\end{align}
where $ \lvert (\cdot) \rvert $ denotes the absolute value of $( \cdot )$ and the axes used in this problem are shown in Fig. \ref{3dgeom}. The Young's modulus and Poisson's ratio used in this problem are
\begin{equation} 
 E =  250.0,  \quad \nu = 0.49999 \text{,} 
\end{equation}
respectively. A mesh with $16\times16\times16$ quadratic elements is used in this example. Fig. \ref{3dplots} plots the distribution of the vertical displacement and the hydrostatic stress using CS elements, CAS1 elements, and CAS2 elements. The viewpoint used in Fig. \ref{3dplots} is the same as the viewpoint used in Fig. \ref{3dgeom}. If we only take a look to the displacement distributions, it may look like the amount of volumetric locking that CS elements have for this problem and mesh resolution is relatively low. However, when looking at the distributions of the hydrostatic stress, we can see that CS elements undergo large-amplitude spurious oscillations while CAS1 elements and CAS2 elements excise these spurious oscillations.

\section{Conclusions and future work}\label{sec6}

In this work, plane strain and three-dimensional linear elasticity are used to investigate how to effectively overcome volumetric locking in quadratic NURBS-based discretizations. We developed two assumed-strain treatments, named CAS1 and CAS2 elements, that vanquish volumetric locking in nearly-incompressible solids. CAS1 elements linearly interpolate the strains at the knots in each direction for the term involving the first Lam\'e parameter in the variational form. CAS2 elements linearly interpolate the dilatational strains at the knots in each direction. For both CAS1 and CAS2 elements, assumed strains with $C^0$ continuity across element boundaries are obtained for a displacement vector with $C^1$ continuity across element boundaries. The effects of volumetric locking are not only smaller displacements than expected, but also large-amplitude spurious oscillations of normal stresses. The spurious oscillations of normal stresses can take place for quite refined meshes for which the displacement values are already accurate. Thus, when the effectivity of a locking treatment is evaluated, it is not enough to only study the accuracy of the displacements. The accuracy of the normal stresses must be studied as well. Both CAS1 and CAS2 eliminate the spurious oscillations of normal stresses. For a given mesh, CAS1 and CAS2 elements result in very similar level of accuracy (when small differences are found, CAS1 elements are frequently slightly more accurate than CAS2 elements). Therefore, the authors favor the use of CAS1 elements since this element type requires fewer operations per quadrature point than CAS2 elements. For both CAS1 and CAS2 elements, the same level of accuracy is obtained with either $2^d$ or $3^d$ Gauss-Legendre quadrature points per element. Thus, $2^d$ Gauss-Legendre quadrature points per element can be used to speed up the simulations. Future research directions include extending the proposed locking treatments to nearly-incompressible hyperelastic solids and elastoplastic solids. As shown in \cite{nguyen2022leveraging}, locking not only negatively affects discretizations of boundary-value problems, but also discretizations of eigenvalue problems. Thus, studying the spectral accuracy of CAS and CS elements for nearly-incompressible linear elasticity is another interesting direction of future work.

\bmhead{Acknowledgments}

H. Casquero and M. Golestanian were partially supported by Ansys Inc. and the NSF grant CMMI-2138187.


%%\bibliography{Bibliography.bib}% common bib file

\end{document}